\documentclass[%
reprint,
superscriptaddress,
 amsmath,amssymb,
 prapplied,
]{revtex4-1}

\usepackage{graphicx}
\usepackage{dcolumn}
\usepackage{bm}
\usepackage{hyperref}
\usepackage{xspace} 
\usepackage{mathrsfs} 
\usepackage[inline]{enumitem} 
\usepackage{color}
\usepackage{siunitx}
\usepackage{mathptmx}
\DeclareMathAlphabet{\pazocal}{OMS}{zplm}{m}{n}

\newcommand{\HH}{\ensuremath{\pazocal{H}}\xspace}
\newcommand{\smmu}{\ensuremath{\mu_{\mathrm{s}}}\xspace}

\newcommand{\smKu}{\ensuremath{k_{\mathrm{u}}}\xspace}

\newcommand{\vampire}{\textsc{vampire}\xspace}

\newcommand{\Mags}{\ensuremath{M_{\mathrm{s}}}\xspace}

\newcommand{\Keff}{\ensuremath{K_{\mathrm{eff}}}\xspace}
\newcommand{\As}{\ensuremath{A_{\mathrm{s}}}\xspace}
\newcommand{\Ku}{\ensuremath{K_{\mathrm{u}}}\xspace}
\newcommand{\EnBar}{\ensuremath{E_{\mathrm{b}}}\xspace}
\newcommand{\StabilityF}{\ensuremath{\Delta}\xspace}

\newcommand{\muzero}{\ensuremath{\mu_0}\xspace}

\begin{document}


\title{Atomistic investigation of the temperature and size dependence\\ of the energy barrier of CoFeB/MgO nanodots}

\author{A.~Meo}
 \email{andrea.m@msu.ac.th, now working at Mahasarakham University (Thailand)}
\affiliation{Department of Physics, University of York, Heslington, York YO10 5DD United Kingdom.}
\author{R~.Chepulskyy}
\affiliation{Samsung Electronics, Semiconductor R\&D Center (Grandis), San Jose, CA 95134, USA}
\author{D~.Apalkov}
\affiliation{Samsung Electronics, Semiconductor R\&D Center (Grandis), San Jose, CA 95134, USA}
\author{R.~W.~Chantrell}
\affiliation{Department of Physics, University of York, Heslington, York YO10 5DD United Kingdom.}
\author{R.~F.~L.~Evans}
\affiliation{Department of Physics, University of York, Heslington, York YO10 5DD United Kingdom.}


\begin{abstract}
The balance between low power consumption and high efficiency in memory devices is a major limiting factor in the development of new technologies. 
Magnetic random access memories (MRAM) based on CoFeB/MgO magnetic tunnel junctions (MTJs) have been proposed as candidates to replace the current technology due to their non-volatility, high thermal stability and efficient operational performance. 
Understanding the size and temperature dependence of the energy barrier and the nature of the transition mechanism across the barrier between stable configurations is a key issue in the development of MRAM. 
Here we use an atomistic spin model to study the energy barrier to reversal in CoFeB/MgO nanodots to determine the effects of size, temperature and external field. We find that for practical device sizes in the 10-50 nm range the energy barrier has a complex behaviour characteristic of a transition from a coherent to domain wall driven reversal process. Such a transition region is not accessible to simple analytical estimates of the energy barrier preventing a unique theoretical calculation of the thermal stability. The atomistic simulations of the energy barrier give good agreement with experimental measurements for similar systems which are at the state of the art and can provide guidance to experiments identifying suitable materials and MTJ stacks with the desired thermal stability.
\end{abstract}

\maketitle

\section{\label{sec:introduction}Introduction}
The balance between low power consumption and high speed in memory devices is one of the most limiting factors in the development of new energy efficient memory technologies. 
Magnetic random access memory (MRAM), a storage device where the data is stored as magnetic state rather than electrical charge, has been proposed as a candidate able to address the power issue because of its non-volatility, while maintaining high performance in writing and reading processes. 
The main component of a MRAM is the magnetic tunnel junction (MTJ), that is a multilayer structure composed, in its simplest design, of two metallic ferromagnets sandwiching a thin non-magnetic insulator.
CoFeB/MgO-based MTJs exhibit high thermal stability, low threshold current and high tunnelling magneto-resistance (TMR) signal and thus they are among the most promising candidates for commercial MRAM. Nevertheless, such technology needs to maintain
these features when scaled below \SI{40}{nm} in the lateral dimension in order to compete with the density of current silicon-based devices. 

Understanding the size and temperature dependence of the energy barrier (\EnBar) and the nature of the transition mechanism across the barrier between stable configurations is a key challenge in the development of MRAM. 
Experimental studies have investigated the energy barrier of magnetic tunnel junctions (MTJs) as function of size \cite{sato2011,sato2012,sato2014,Sun2013,takeuchi2015,Gajek2012a}. \citet{Gajek2012a} determine the energy barrier via measurements of the switching current as a function of the switching frequency and find that the energy barrier scales quadratically with the diameter of the device as expected for a macrospin.
A sharp change in the size dependence of the energy barrier resembling a linear trend for diameters larger than \SI{30}{nm} can be observed in Fig.~3 of the same work and the average coercivity of the whole MTJ junction flattens for larger dimensions.
\citet{sato2011,sato2012,sato2014,Sun2013,takeuchi2015} found values lower than expected from a macrospin model for a system larger than the estimated single domain size, which suggests domain nucleation as the reversal mechanism.
In a recent work \citet{Enobio2018} extract the energy barrier for MTJs similar to those investigated by \cite{takeuchi2015} via retention time measurements.
They conclude that the crossing over the energy barrier might be described by a magnetic reversal mechanism different from nucleation. According to their reasoning the energy barrier is independent of the junction diameter in case of nucleation and the fact that they find a dependence on the MTJ diameter calls for some different explanation.
The theoretical analysis performed so far on similar systems  \cite{Munira2015a,chaves2013,chaves2015} are based on zero temperature micromagnetic modelling. 
The continuum approach on which standard micromagnetism is developed begins to fail with the miniaturisation of devices down to a few nanometres due to the inability to describe elevated temperatures, surface and interfacial effects and complex magnetic ordering \cite{CuadradoPRApp2018}. 
Furthermore, the effect of temperature has been experimentally investigated only by \citet{takeuchi2015}. 

Here we study the energy barrier to magnetisation reversal in CoFeB/MgO nanodots focusing on its size, temperature and field dependence using an atomistic spin model. 
The results show a dependence of the energy barrier for a given thickness and temperature which scales quadratically with the diameter for systems smaller than the estimated single domain size, whereas the trend becomes linear for larger dimensions. The former behaviour can be explained by a macrospin and the coherent reversal mechanism, while the latter is characteristic of a domain wall mediated switching process. 

\section{Methods}
\label{sec:methods}
We perform simulations based on the atomistic spin model as implemented in the \vampire software package \cite{vampire} with a localised Heisenberg approximation for the exchange: 
\begin{equation}
\HH = - \sum_{i < j} J_{ij} \vec{S}_i\cdot \vec{S}_j - \sum_{i} \smKu^i (\vec{S}_i\cdot \hat{e})^2 - \sum_i \smmu^i \vec{S}_i \cdot \vec{\mathrm{B}}_{\mathrm{app}} + \HH_{\mathrm{dmg}} .
  \label{eq:Hamiltonian}
\end{equation}
$J_{ij}$ is the exchange coupling constant for the interaction between the spins on site $i$ and $j$, $\smKu^i$ is the uniaxial energy constant on site $i$ along the easy-axis $\hat{e}$, $\smmu^i$ is the atomic spin moment on the atomic site $i$ and $\vec{\mathrm{H}}_{\mathrm{app}}$ is the external applied field.  
The magnetostatic contributions $\HH_{\mathrm{dmg}}$ to the energy of the system are calculated using a modified macrocell approach where the contribution within each cell is taken into account following the approach proposed by Bowden \cite{bowden}, explicitly computing the interaction tensor from the atomistic coordinates.
To obtain the energy barrier and its temperature dependence we use the constrained Monte Carlo (cMC) algorithm, a modified Monte Carlo (MC) algorithm \cite{CMC_vampire} with an adaptive spin update algorithm \cite{AlzateCardonaJPCM2019}. 
A standard MC algorithm allows the determination of the magnetic properties at thermal equilibrium. 
In such a condition we cannot access the magnetic anisotropy since the magnetisation aligns along the equilibrium direction, generally the easy-axis.
To circumvent this, one can keep the system in a quasi equilibrium state.
Such an approach has been exploited in the cMC method, which acts on two spins simultaneously and allows the direction of the global magnetisation to be constrained during the simulation along specific directions, whilst allowing individual spins $\vec{S}_i$ to reach thermal equilibrium.
Since the system is not in equilibrium, the total internal torque $\vec{\tau}$ acting on the magnetisation $\vec{M}$ \cite{CMC_vampire} does not vanish.
For a system at constant temperature the magnitude of the torque acting on the system, whose magnitude represents the work done on the system, can be expressed as:
\begin{equation}
\vec{\tau}  = -\frac{\partial \pazocal{F}}{\partial\vartheta},
\end{equation}
where $\vartheta$ is the constraining direction and $\pazocal{F}(\vec{\mathrm{M}})$ is the Helmholtz free energy of the system which measures the amount of work that can be obtained in a physical system at constant temperature and volume.
We can then compute the anisotropy energy as the variation of the free energy:
\begin{equation}
\Delta \pazocal{F} = -\int d\vartheta \vec{\tau}.
\end{equation}
Calculating $\Delta \pazocal{F}$ at different temperatures allows the reconstruction of the temperature dependence of the effective magnetic anisotropy.

We model CoFeB/MgO nanodots, the constituent of a MTJ, as cylindrical alloy films with a body-centred cubic (\textsc{bcc}) crystal structure.
At the CoFeB/MgO interface the system is characterised by large perpendicular uniaxial anisotropy induced by the interfacial MgO \cite{Yang2011,Turek2003} whose value is derived from the temperature dependence of the macroscopic anisotropy energy density for CoFeB/MgO thin films \cite{Sato2018}, whereas the rest of the material is assumed to have zero anisotropy. 
The atomistic exchange parameters are obtained from a mean field approach with a correction for spin-wave excitations \citep{vampire-rev,Garanin1996} to reproduce the experimental Curie temperature \cite{Sato2018}.

\section{Results}
\label{sec:discussion}
Here we focus on the size and temperature dependence of the energy barrier and the mechanism of the transition over the barrier in CoFeB/MgO nanodots of thickness \SI{1}{nm}.
Moreover, we have determined the role of an external magnetic field on the energy barrier and investigated the reversal mechanism.

\subsection{Thermal stability in zero field}
\label{subsec:Ebar}
\begin{figure}[tbp]
\centering
\includegraphics[angle = 0,width = 0.9\columnwidth]{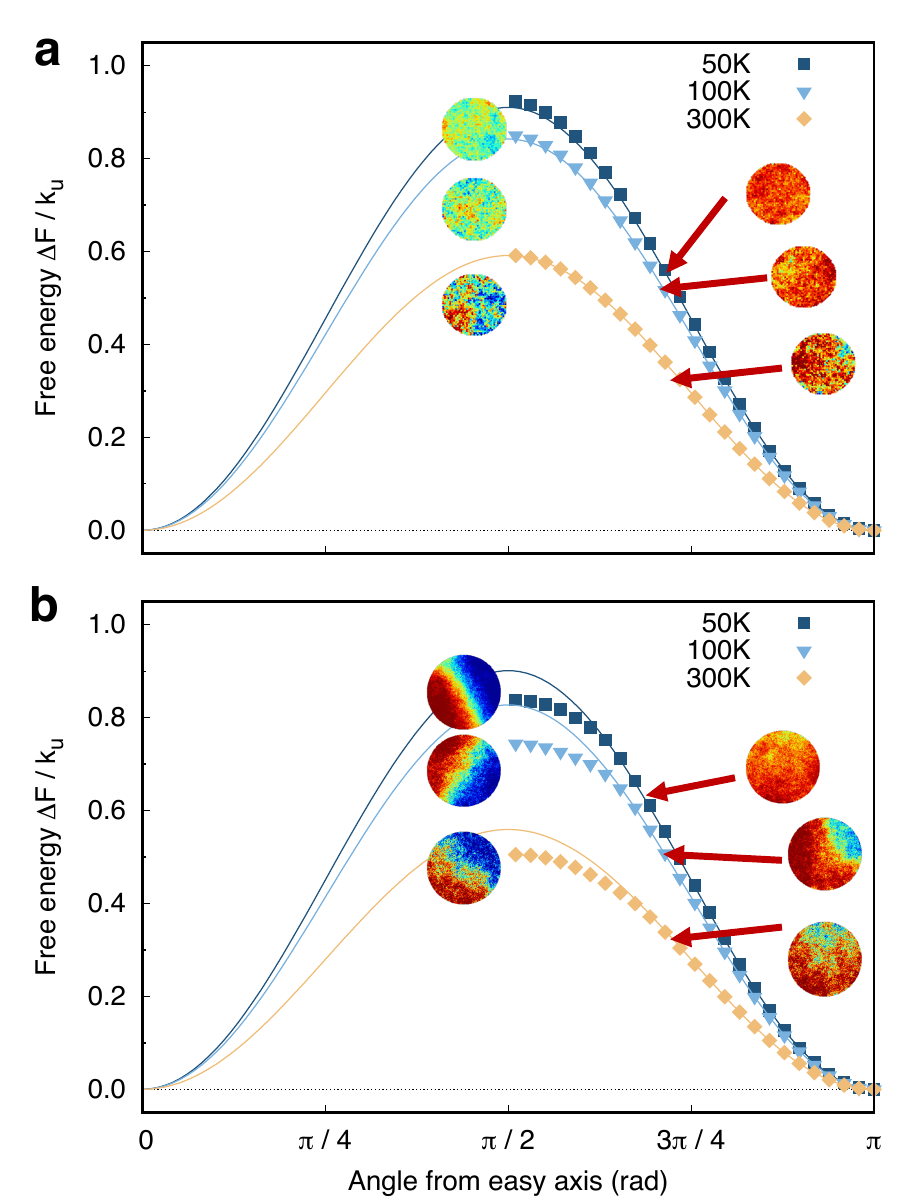}
\caption{Angular dependence of scaled the energy barrier scaled by the uniaxial energy constant at \SI{0}{K} for a \SI{10}{nm} (a) and \SI{30}{nm} (b) dot.  The insets show snapshots of the out-of-plane component of the magnetisation (red = spin-down, green = in-plane, blue = spin-up) at \SIlist{50;100;300}{K} for constraint angles of the magnetisation of $\pi/2$ and $2/3\pi$. Lines are the fit of the data in the region $[4/5 \pi: \pi ]$ with a $\sin(\vartheta)^2$ function. 
}
\label{fig:Ebar_vs_th}
\end{figure}
We use the cMC algorithm, described in the methods section, to calculate the angular dependence of the restoring torque via the constraint of the total magnetisation away from the easy-axis direction at different temperatures and diameters.
Using the cMC method we compute the total torque acting on the magnetisation and by integrating this over the angular distribution we obtain the effective energy barrier separating the two stable states of the system. 
Fig.~\ref{fig:Ebar_vs_th} presents the angular dependence of the free energy for diameters \SIlist{10;30}{nm} at different temperatures. 
The \SI{10}{nm} dot closely follows the $\sin(\vartheta)^2$ behaviour characteristic of a uniaxial system and coherent reversal, where $\vartheta$ is the angle formed by magnetisation and easy axis. 
Snapshots of the out-of-plane spin configuration, shown in the inset of Fig.~\ref{fig:Ebar_vs_th}(a), confirm the coherent nature of the reversal, even though thermal effects cause large fluctuations at small system dimensions and the switching is not completely coherent.
The free energy of the \SI{30}{nm} disc deviates from the above trend decreasing as angles approach $\pi/2$. 
This is consistent with a nucleation-type reversal and results in a lower energy barrier than for a coherent mechanism, which is often assumed in MTJ devices, and is confirmed by the spin configurations in Fig.~\ref{fig:Ebar_vs_th}(b). This decrease of the energy barrier in case of nucleation poses issues for technological applications as it yields a lower thermal stability than predicted using a macrospin model and also induces an intrinsic stochastic character to the reversal as the wall velocity during reversal can be reduced due to  pinning sites.

\begin{figure}[tb!]
\centering
\includegraphics[angle = 0,width = 0.9\columnwidth]{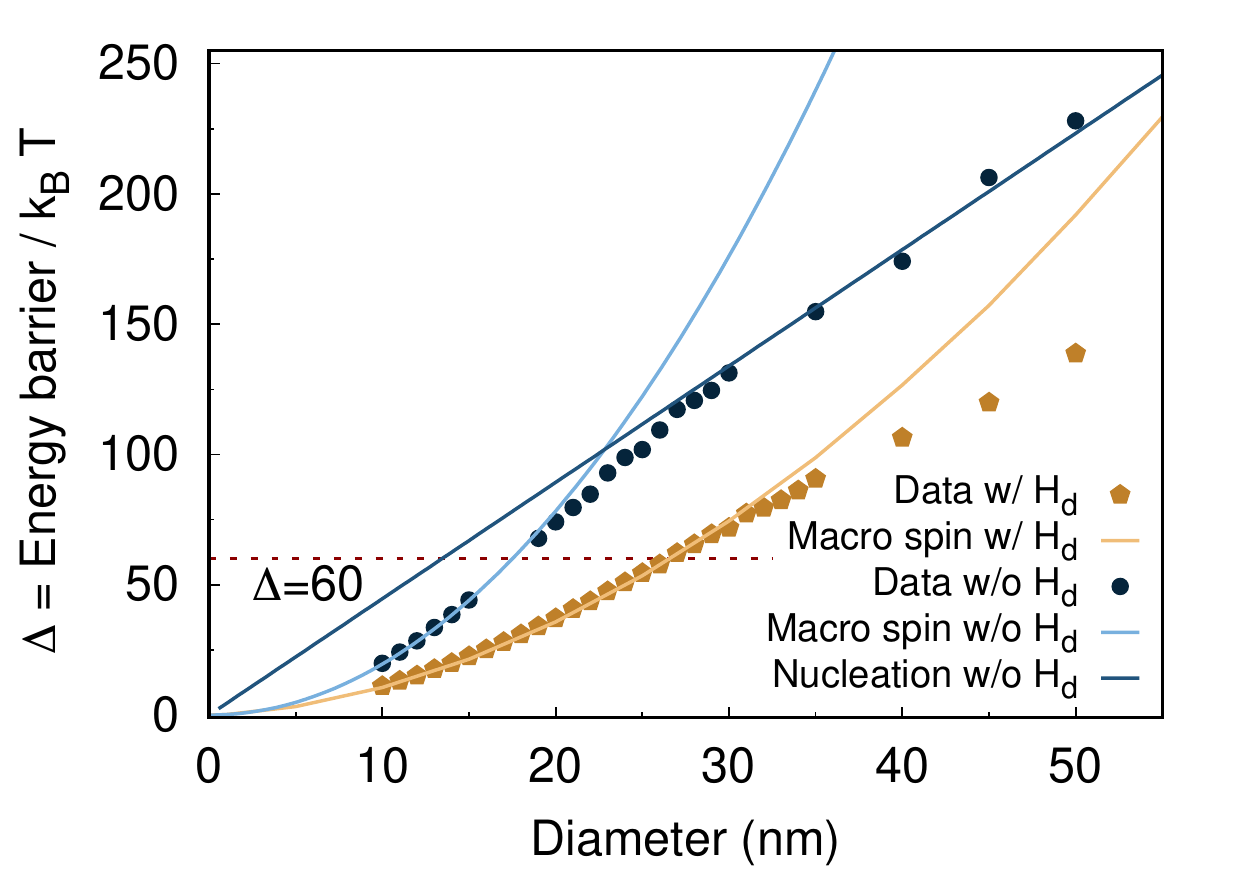}
\caption{Stability factor (Energy barrier/$k_\mathrm{B}T$) as function of diameter for CoFeB/MgO dots at \SI{300}{K}. Black dots and blue line represent data and fit for simulations without magnetostatic interactions, orange diamonds and yellow line describe data and fit for simulations with magnetostatic interactions.
 }
\label{fig:Ebar_vs_d}
\end{figure}
One of the most relevant parameters for applications is the stability factor \StabilityF, defined as the energy barrier normalised by the thermal energy $k_{\mathrm{B}}T$, where $k_{\mathrm{B}}$ is Boltzmann constant and $T$ the absolute temperature. For technological applications such as storage devices, a stability factor larger than 60 at room temperature is required in order to guarantee a data retention of at least ten years. 
In Fig.~\ref{fig:Ebar_vs_d} (yellow line and orange diamonds) we show the size dependence at room temperature of \StabilityF.
\StabilityF is quadratic for dots smaller than \SI{20}{nm}, whereas it starts deviating towards a linear trend for larger sizes. 
The existence of different regimes can be understood in terms of the reversal mechanism of the magnetisation: if the reversal is coherent, \EnBar follows the macrospin behaviour and is given by the analytic expression $\EnBar = \Keff V$ \cite{Skomski}, where \Keff is the effective magnetocrystalline energy density and $V = \pi t d^2 /4$ is the disc volume.
In case of nucleation \EnBar can be obtained from the energy of a domain wall in the centre of the disc $\EnBar = \sigma w$, where $\sigma = 4 \sqrt{\As /\Keff}$ is the domain wall surface energy density, \As the exchange stiffness and $w = d t$ is the surface of the disc of the disc \cite{Skomski}. 
The former yields a quadratic scaling of \EnBar with the diameter, whereas the latter linear, in agreement with the trend of the data.
\Keff includes both the magnetocrystalline anisotropy \Ku and the shape contribution arising from the long-range dipole-dipole interaction among the spins. 
For a uniformly magnetised cylinder the magnetostatic contribution can be written in terms of the demagnetisation tensor $N$:
\begin{equation} 
\Keff = \Ku - \frac{1}{2}\muzero \Mags^2  \frac{( N_{\mathrm{zz}} - 1 )}{2}.
\label{eq:Keff_MS}
\end{equation}
Here \Mags is the saturation magnetisation and $N_{\mathrm{zz}}$ is the $zz$ component of the demagnetisation tensor. 
The second term on the RHS of equation(\ref{eq:Keff_MS}) is the demagnetising energy for a cylinder which is magnetised along the easy axis direction $z$ and $( N_{\mathrm{zz}} - 1 )$ comes from  $(N_{\mathrm{xx}}+N_{\mathrm{yy}}+N_{\mathrm{zz}})=1$ in \emph{SI} units. 
As the demagnetisation energy favours magnetisation alignment along the largest dimension, this contribution yields a smaller anisotropy energy for thin cylinders causing a broadening of the domain wall width and a reduction in the energy barrier, compared with a case where this term is neglected. 
The same expression should not be used for non-uniform magnetisation configurations since the demagnetisation tensor is a macroscopic quantity defined for a uniformly magnetised system.
If we consider a system with an infinitesimal wall in the centre of the system separating two magnetic domains of equal volume, its magnetostatic contribution should be zero. 
However, we cannot properly compute the magnetostatic energy of this magnetic configuration using equation (\ref{eq:Keff_MS}).
Since an analytic formulation for non-uniform magnetised systems is not easily accessible, we compute the size and temperature dependence of \EnBar neglecting the magnetostatic interaction as well, so that $\Keff=\Ku$.

We compare our data obtained with and without the inclusion of magnetostatic interactions with the analytic expression for a macrospin system, where we derive the parameters \As ($\sim \SI{20e-12}{Jm^{-1}}$) and \Ku ($\sim \SI{1e6}{Jm^{-3}}$) from atomistic values, following Ref.\cite{Moreno2016}.  
Excellent agreement between the simulated data and the analytic expression for a coherent reversal is found for diameters smaller than \SIlist{30;20}{nm} for calculations with and without magnetostatic interactions, respectively. 
For larger diameters the data deviate from the above trends and seem to lie in an intermediate regime where there is no available analytic expression.
As the diameter is increased further, the data is well fitted by a linear trend as described by the nucleation theory for simulations where $\Keff=\Ku$.
For simulations where the magnetostatic contribution is included, we observe a similar trend. However, a direct comparison with theoretical expressions is not possible due to the complex angle variation of the total magnetostatic field during reversal. 
A similar analysis has been performed by \citet{chaves2015} by means of a micromagnetic approach at zero temperature and by rescaling the MTJ parameters and size to that of a permalloy disc.
The results of \citet{chaves2015} are in good agreement with the analytic expressions for both the macrospin and nucleation regime.
We point out that in our opinion using the demagnetisation coefficients to determine \Keff for systems with non-uniform magnetisation configurations is not appropriate and leads to an overestimate of the domain wall energy. 
We attribute the good agreement between the data and the theory in this regime to a combination of the scaling of the magnetic properties of the system and micromagnetic simulations.

The technological requirement for memory and storage devices is the retention of data, i.e. the magnetic state, for a minimum of ten years.
This corresponds to a $\StabilityF \geq 60$ at room temperature, underlined by the red dashed line in Fig.~\ref{fig:Ebar_vs_d}.
In our case the smallest element size able to yield the desired thermal stability is around \SI{28}{nm} when we include the shape anisotropy contribution; hence a smaller system would not satisfy the thermal stability requirements.
A solution is to increase the complexity of the stack, as in MTJs with the free layer composed of a double MgO \citep{sato2014,Liu2018} barrier.
Such a design helps in reducing the effect of the stray field and is characterised by a larger interfacial anisotropy due to the increased number of CoFeB/MgO interfaces.
Another alternative is to fabricate MTJs with an elongated free layer to exploit the shape anisotropy of a rod-like system as an additional source of perpendicular anisotropy, which would allow lateral device dimensions below \SI{10}{nm} \cite{PerrissinRSC2018,Watanabe2018,PerrissinJPhysD2019}.
Nevertheless, further studies on both the equilibrium and dynamic properties of such a stack need to be performed to assess the viability of this solution.

\subsection{Effect of an applied field}
\label{subsec:Ebar_field}
\begin{figure}[tpb]
\centering
\includegraphics[angle = 0,width = 0.9\columnwidth]{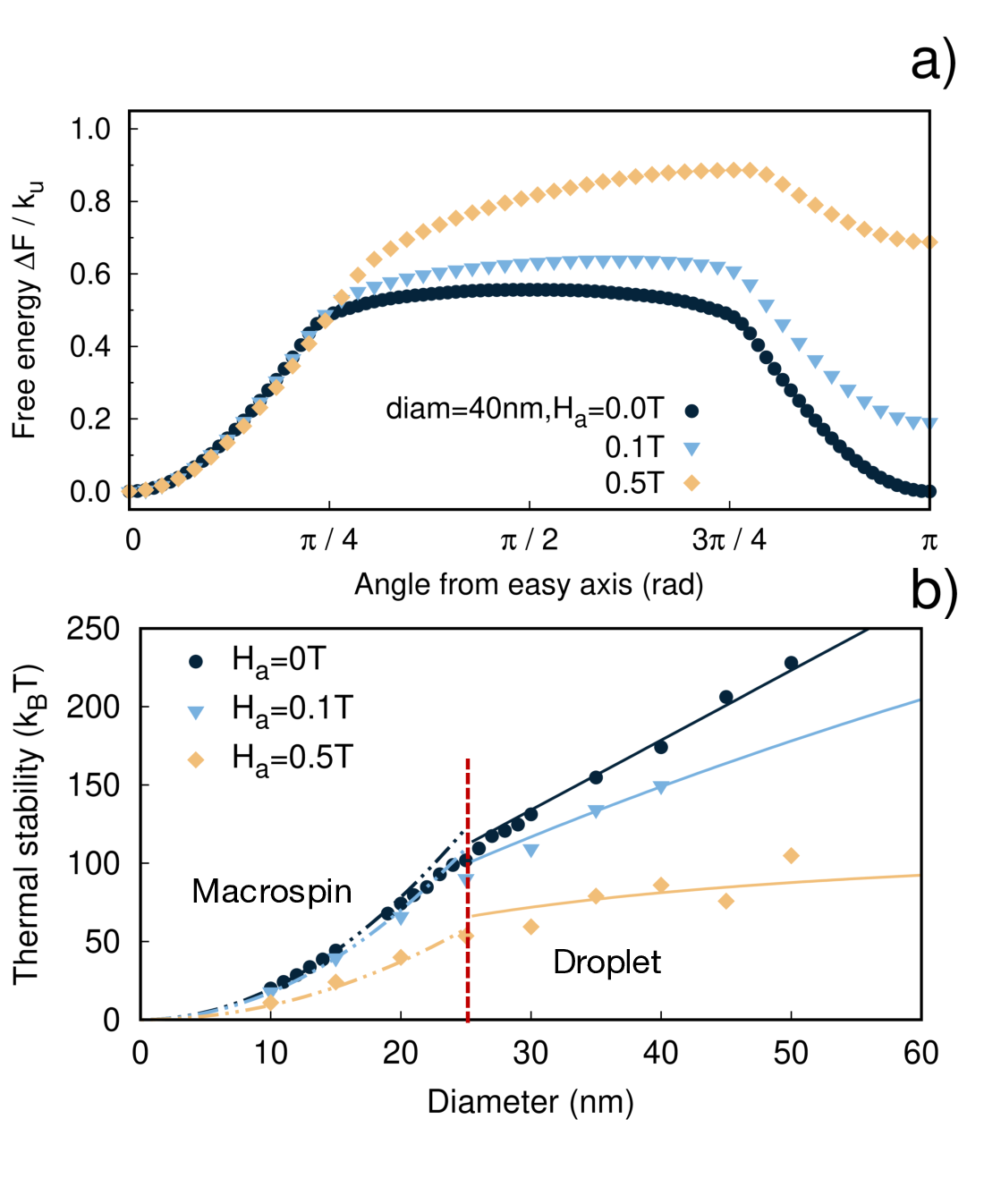}
\caption{
(a) Angular dependence of the energy barrier normalised by the maximum torque for CoFeB/MgO dots of diameter \SI{40}{nm} at \SI{300}{\kelvin} for $H_{\mathrm{a}}=$\SIlist{0.0;0.1;0.5}{\tesla} applied along the positive z-direction.
(b) Thermal stability (Energy barrier/$k_\mathrm{B}T$) as function of diameter for CoFeB/MgO dots at \SI{300}{\kelvin} for an external field as in (a). 
Points represent the data, solid lines represent the the analytic model describing the droplet theory and dotted lines describe the macrospin model for uniform magnetisation for dimensions smaller than single domain size. 
}
\label{fig:Ebar_vs_d_field}
\end{figure}
An applied field acting on the reference layer of a MTJ can alter the energy landscape of this layer, for instance raising one minimum and lowering the other if the field is perpendicular to the stack, and affect the reversal mechanism \cite{Meo2017a}. 
For simple MTJ geometries such as a single free layer MTJ \citep{ikeda2010,sato2014,Enobio2018}, the recording layer is subjected to the stray field coming from the reference layer which can affect the stability of the system \cite{Bapna2016,Jenkins2019}.
We perform simulations at \SI{300}{K} applying an external field $B_{\mathrm{a}}=$\SIlist{0.0;0.1;0.5}{\tesla} along the positive $z$-direction perpendicular to the dot.
Fig.~\ref{fig:Ebar_vs_d_field}(a) depicts the free energy as function of the angle between the total magnetisation and the easy axis for dots of diameter \SI{40}{nm} at $T=$ \SI{300}{K} for the different $B_{\mathrm{a}}$.
The effect of the external field is to decrease the energy of the minimum corresponding to magnetisation aligned with the external field and to rise the other stable state.
As a consequence, the energy barrier between the two stable configurations becomes nonequivalent. 
Similarly to the zero field case, a flattening of the energy barrier at large angle can be observed. 
However, the application of an external field applied along the stable direction of the magnetisation aids the nucleation and  allows the nucleation reversal mechanism at smaller angles. 
We fit the angular dependence of the energy barrier by adding $-2 B_{\mathrm{a}} \cos(\vartheta_0-\vartheta)$ to $\Ku \sin^2(\vartheta)$, where $\vartheta_0$ is the angle between the easy axis and the applied field. 
Since the field is applied along the easy axis direction, we set $\vartheta_0=0$. 
For small diameters the free energy follows the expected analytic expression for the total energy, as expected for coherent rotation. For larger diameters the system is characterised by a non-uniform reversal mode and the analytic expression cannot reproduce the data. 

We compute the energy barrier integrating the torque over the angular dependence and plot the stability factor $\StabilityF$ as a function of particle diameter and applied field at \SI{300}{K} in Fig.~\ref{fig:Ebar_vs_d_field}(b). 
For diameters less than \SI{25}{nm}, close to the single domain size of the system, $\StabilityF$ follows a quadratic behaviour well fit by a macrospin model including the effect of the external field mentioned above, as demonstrated by the dotted lines.
Deviations from this behaviour occur as the system size approaches the limiting single domain size and for larger diameters $\StabilityF$ depends linearly on the diameter of the disk, with a slope that decreases for increasing applied fields.
In this regime we cannot rely on the macrospin theory, as already seen in the zero field case.
The droplet theory \cite{Hinzke1998,Hinzke1999,Nowak1999} provides an analytic approach to account for the effect of an external field on the energy barrier when the transition between the minimum energy states is non-uniform. If magnetostatic fields are neglected, we can write: 
\begin{figure}[tpb]
\centering
\includegraphics[angle = 0,width = 0.6 \columnwidth]{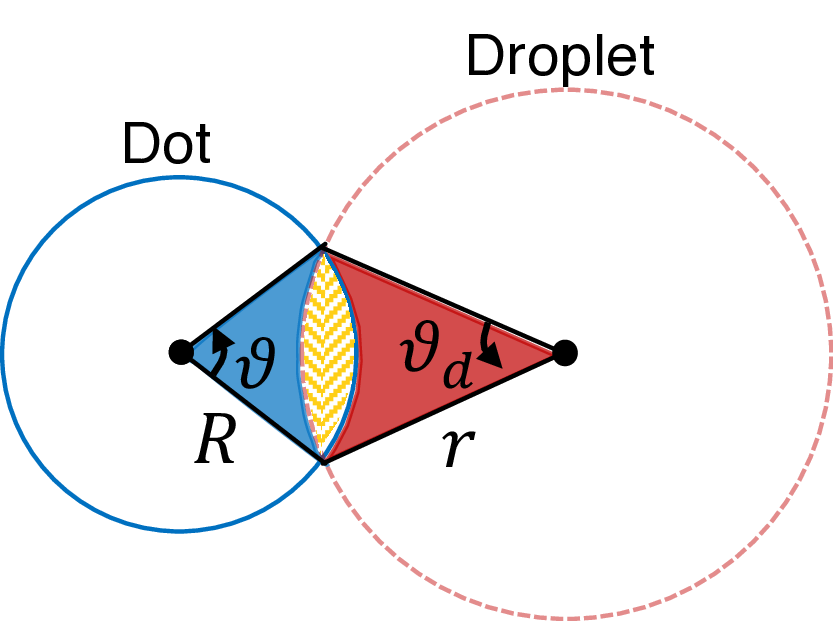}
\caption{Sketch showing the parameters used in the calculation of the energy barrier in the droplet model.
The reversed magnetised domain is described by the yellow dashed region. 
The intersection between the blue and red circular sectors within the dot and droplet respectively determines the position and size of the nucleated area.
$t \vartheta_{\mathrm{d}} r$ is the contact area between the two domains and $\{ t R^2 \left[\vartheta - \sin(\vartheta)\right] + r^2 \left[ \vartheta_{\mathrm{d}} - \sin(\vartheta_{\mathrm{d}}) \right] \} /2$ is the volume of the nucleated domain.
 }
\label{fig:droplet}
\end{figure}
\begin{align}
\label{eq:droplet}
\EnBar ^{\mathrm{drop}} (R) = &  
\sigma t \vartheta_{\mathrm{d}} r -\\ 
& 2 \Mags B_{\mathrm{a}} t \frac{R^2 \left[\vartheta - \sin(\vartheta)\right] + r^2 \left[ \vartheta_{\mathrm{d}} - \sin(\vartheta_{\mathrm{d}}) \right] }{2} , \nonumber
\end{align}
where $R$, $r$, $\vartheta$, $\vartheta_{\mathrm{d}}$ refer to Fig.~\ref{fig:droplet}, $\sigma$ is the domain wall energy assuming a domain wall at the centre of the system, \Mags the saturation magnetisation of the system and $B_{\mathrm{a}}$ is the magnitude of the external applied field.
One can easily prove that if $B_{\mathrm{a}}=\SI{0}{T}$  equation~\ref{eq:droplet} reduces to the nucleation model.
We fit our data for diameters larger than \SI{25}{nm} with  equation~\ref{eq:droplet}, represented by the solid lines in Fig.~\ref{fig:Ebar_vs_d_field}(b).
The model predicts an asymptotic behaviour of the energy barrier as a function of increasing diameter for a sufficiently strong magnitude of the field. 
We can start observing this for  the external field  in agreement with the trend shown by our results for $B_{\mathrm{a}}=\SI{0}{T}$, although larger diameters would be required to reach the asymptotic behaviour.
In a similar study, \citet{chaves2015} investigate the effect of an external field on the stability of the disc, such as the stray field acting on the free layer of a MTJ coming from the reference layer by means of a zero temperature micromagnetic approach. 
\citet{chaves2015} predict a saturation of the energy barrier for large diameters, approaching \SI{100}{nm} or larger, and strong applied fields in agreement with the droplet model.
We stress that we did not include the magnetostatic contribution in these simulations to allow a direct comparison with the theoretical droplet model, as it is not clear how to analytically account for this term in the droplet theory formalism.
Moreover, we argue that the use of demagnetisation tensors when dealing with non-uniform magnetisation configurations is not appropriate as this is a macroscopic quantity defined in the case of uniform magnetic system.
 
\subsection{Comparison with experimental results}
\label{subsec:Ebar_comparison}
\begin{figure}[tpb]
\centering
\includegraphics[angle = 0,width = 0.9\columnwidth]{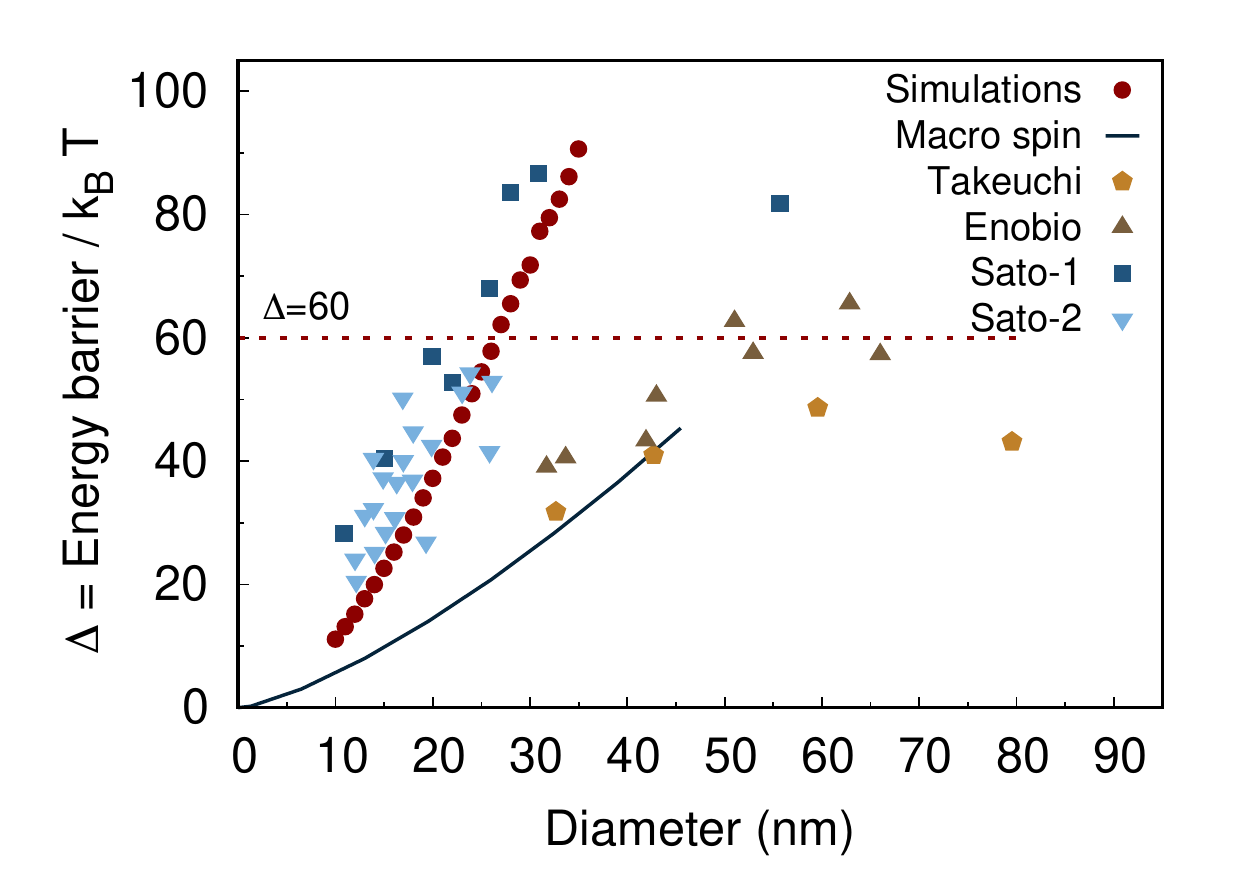}
\caption{Comparison between simulated (red dots) stability factor (Energy barrier/$k_\mathrm{B}T$) as function of diameter for CoFeB/MgO dots at \SI{300}{K} and experimental results from references \citep{sato2014,takeuchi2015,Enobio2018}. Sato-1 (blue squares) and Sato-2 (light-blue downwards triangle) refer to series1 and series2 of \cite{sato2014}, respectively.
 }
\label{fig:comparison_Ebar_vs_d}
\end{figure}
Finally we compare our simulation results against experiments performed by \citet{sato2014,takeuchi2015,Enobio2018} in Fig.~\ref{fig:comparison_Ebar_vs_d}.
The simulated data is described by red dots, blue squares and light-blue downwards triangle refer to series1 and series2 of Ref. \cite{sato2014}, respectively. 
Orange diamonds and brown upwards triangles are extracted from the works of \citet{takeuchi2015,Enobio2018}, respectively. 
The latter two investigate MTJs composed of a single CoFeB/MgO free layer with $\Keff \sim \SI{1.9e5}{Jm^{-3}}$, $\Mags \sim \SI{1.3}{T} $ and $\As \sim \SI{30e-11}{Jm^{-1}} $. 
\citet{sato2014} study MTJs with MgO/CoFeB/Ta/CoFeB/MgO recording layer structure characterised by $\Keff \sim \SI{9.4e4}{Jm^{-3}}$, $\Mags \sim \SI{1}{T} $ and $\As \sim \SI{20e-11}{Jm^{-1}}$ and as a result are more stable.

We can see that our simulations agree with the data from Ref. \cite{sato2014} up to diameters of \SI{35}{nm}, whilst the stability factor results are much larger than those reported by Ref. \cite{takeuchi2015} and \cite{Enobio2018}.
\citet{takeuchi2015} and \citet{Enobio2018} studied MTJs with lower perpendicular anisotropy than in our simulations, which can explain our larger energy barriers.
We parametrise the macrospin model for \EnBar using our atomistic parameters and we estimate the size dependence of $\StabilityF$ for a dot of thickness \SI{1.3}{nm} within the expected uniform reversal region, shown by the solid line in Fig.~\ref{fig:comparison_Ebar_vs_d}. 
Our prediction seems to agree with the experimental data in Ref. \citep{takeuchi2015,Enobio2018} for diameters smaller than the critical domain size, estimated around \SI{45}{nm}. 
However, we need calculations to prove the agreement, calculations that become particularly intensive at large dimensions.

The incidental agreement between experiments performed on MTJs with a double MgO layer \cite{sato2014} and our simulations is unexpected since we simulate a single CoFeB/MgO free layer and the two structures are not comparable in properties. 
We can speculate that the small thickness used in our simulations, \SI{1}{nm}, yields a $\Keff$ similar to the double stack studied in the experiments.
However, to assess more realistically an agreement we would need to perform energy barrier calculations on an analogous system.
We also point out that in our simulations we do not account for structural defects and fabrication damage, something that occurs in real devices.

Finally, we can observe as \citet{sato2014} and \citet{takeuchi2015} obtain a value of \StabilityF which suggests a constant \EnBar for large diameters, rather far from the single domain size of their systems, which they associate to nucleation mechanisms.
While we agree on the nature of the reversal mechanism for such dimensions, we believe that \StabilityF should still show a dependence on the diameter of the dots, even for large dimensions. 
In both these works \EnBar is obtained by measuring the switching probability as a function of the magnitude of the applied magnetic field and such fields might affect \EnBar. 
Another possible reason for the constant trend could be the presence of fringe fields due to the reference layer of the MTJ \cite{Jenkins2019, Bapna2016} that could reduce \EnBar, as also discussed in the analysis of the effect of an external field.

Overall our results demonstrate the ability of an atomistic spin model to calculate the energy barrier for technologically relevant sizes of realistic MTJs at pertinent temperatures yielding values close to experiments, even in the nucleation regime where the magnetostatic contribution needs to be accurately accounted for. 

\section{Conclusions}
The energy barrier is the key parameter in the thermal stability of magnetic tunnel junctions, the main component of MRAM devices, and needs to be characterised.
To deal with the devices at the nanoscale we have used an atomistic spin model to investigate the size, temperature and field dependence of the energy barrier in ultra thin CoFeB/MgO discs comprising the free layers of magnetic tunnel junctions.
We have found that a transition from coherent to domain wall mediated reversal occurs around the critical domain size around \SI{30}{nm}.
These two regimes are not separated by a sharp transition and the intermediate region is not captured by analytic descriptions.
Therefore it is important to have approaches that provide understanding and are able to accurately characterise the thermal stability at dimensions that are of high technological interest. 
We have also investigated the impact of external fields showing that applied fields can reduce the energy barrier and affect the reversal mechanism, as in the case of fringing fields. 
These effects should be taken into account in real device design as their effects could be detrimental for the device performance.
To conclude, we have shown that the atomistic spin model parameterised using realistic values can extract and characterise the energy barrier in systems that are at the state of the art and can provide guidance to experiments identifying suitable materials and MTJ stacks with the desired thermal stability. 

\section*{acknowledgements}
The authors gratefully acknowledge funding from the Samsung SGMI programme. 

\section*{Author contributions}
AM performed the atomistic simulations, analysed the results and drafted the paper. AM, RFLE and RWC conceived and designed the study. All authors contributed to the writing of the paper.

%

\end{document}